\documentclass[11pt,dvips]{article}

\usepackage{epsfig,times} 
\usepackage{wrapfig}
\makeatletter
\renewcommand{\section}{\@startsection{section}{1}{0in}
	{0.4\baselineskip}{0.1\baselineskip}{\Large\bf}}
\renewcommand{\subsection}{\@startsection{subsection}{2}{0in}
	{0.25\baselineskip}{-\baselineskip}{\large\bf}}
\renewcommand{\subsubsection}{\@startsection{subsubsection}{3}{0in}
	{0.1\baselineskip}{-\baselineskip}{\normalsize\bf}}

\makeatother
\begin{document}

%
\thispagestyle{myheadings}
%
\markright{OG 2.1.03}
\begin{center}
%
{\LARGE \bf Hour-Scale Multiwavelength Variability in Markarian 421}
\end{center}

\begin{center}
%
%
{\bf M.\,Catanese$^1$, I.H.\,Bond$^2$, S.M.\,Bradbury$^2$, A.C.\,Breslin$^3$, 
J.H.\,Buckley$^4$, A.M.\,Burdett$^{2,5}$, D.A.\,Carter-Lewis$^1$, 
M.F.\,Cawley$^6$, S.\,Dunlea$^3$, M.\,D'Vali$^2$, 
D.J.\,Fegan$^3$, S.J.\,Fegan$^5$, J.P.\,Finley$^7$, J.A.\,Gaidos$^7$, 
T.A.\,Hall$^7$, A.M.\,Hillas$^2$, D.\,Horan$^3$, J.\,Knapp$^2$, 
F.\,Krennrich$^1$, S.\,Le Bohec$^1$, R.W.\,Lessard$^7$, C.\,Masterson$^3$, 
P.\,Moriarty$^8$, J.\,Quinn$^3$, H.J.\,Rose$^2$, F.W.\,Samuelson$^1$, 
G.H.\,Sembroski$^7$, V.V.\,Vassiliev$^5$, T.C.\,Weekes$^5$, 
L.\,Maraschi$^9$, G.\,Fossati$^{10}$, F.\,Tavecchio$^9$, L.\,Chiappetti$^{11}$,
A.\,Celotti$^{10}$, G.\,Ghisellini$^9$, P.\,Grandi$^{12}$, E.\,Pian$^{13}$, 
G.\,Tagliaferri$^9$, A.\,Treves$^{14}$\\
}

{\it
$^1$Department of Physics and Astronomy, Iowa State University, Ames, IA 50011, USA\\
$^2$Department of Physics, Leeds University, Leeds, LS2 9JT, UK\\
$^3$Experimental Physics Department, University College, Belfield, Dublin 4, Ireland\\
$^4$Department of Physics, Washington University, St. Louis, MO 63130, USA\\
$^5$Fred Lawrence Whipple Observatory, Harvard-Smithsonian CfA, Amado, AZ 85645, USA\\
$^6$Physics Department, St. Patrick's College, Maynooth, County Kildare,
	Ireland\\
$^7$Department of Physics, Purdue University, West Lafayette, IN 47907, USA\\
$^8$Department of Physical Sciences, Galway-Mayo IT, Ireland\\
$^9$Osservatorio Astronomico di Brera, Milano, Italy\\
$^{10}$SISSA/ISAS, Trieste, Italy\\
$^{11}$IFC/CNR, Milano, Italy\\
$^{12}$IAS/CNA, Roma, Italy\\
$^{13}$TESRE/CNR, Bologna, Italy\\
$^{14}$Universita dell'Insubria, Como, Italy\\
}
\end{center}

\begin{center}
{\large \bf Abstract\\} 
\end{center}
Markarian\,421 was observed for about four days with {\it Beppo}SAX and the
Whipple Observatory $\gamma$-ray telescope in April 1998. A
pronounced, well-defined, flare with hour-scale variability was
observed simultaneously in X-rays and very high energy $\gamma$-rays.
These data provide the first evidence that the X-ray and TeV
intensities are well correlated on time-scales of hours.  While the
rise of the flare occurred on a similar time-scale in the two
wavebands, the decay of the flare was much more rapid in
$\gamma$-rays, providing the first clear indication that the X-ray and
$\gamma$-ray emission may not be completely correlated in Markarian\,421.
\vspace{-0.5ex}
%

\vspace{1ex}

%
%
\section{Introduction:}
\label{intro.sec}

Blazars are a class of active galactic nuclei whose emission is
believed to arise predominantly from a relativistic jet whose axis
makes a small angle with our line of sight.  More than 60 blazars have
been detected with the Energetic Gamma-Ray Experiment (EGRET) (Hartman
et al.\ 1999).  A few BL Lacertae objects (BL Lacs), a sub-class of
blazars, have also been detected as TeV $\gamma$-ray emitters (Ong
1998).  No model for the origin of the $\gamma$-ray emission is
generally accepted at this time.  Two popular classes are those in
which high energy electrons produce the $\gamma$-rays by inverse
Compton scattering of low energy photons (e.g., Maraschi, Ghisellini
\& Celotti 1992; Dermer, Schlickeiser \& Mastichiadis 1992; Sikora,
Begelman \& Rees 1994) and those in which high energy protons produce
$\gamma$-rays by initiating cascades in the jets (e.g., Mannheim
1993).  Contemporaneous observations at several wavelengths can be
used to derive physical conditions in and around the blazar jet and
may resolve which emission mechanism operates in the objects.

Markarian\,421 (Mrk\,421) is the closest known BL Lac ($z=0.031$) and
is an established very high energy (VHE, E$>$250\,GeV) $\gamma$-ray
source (e.g., Punch et al.\ 1992; Petry et al.\ 1996).  Mrk\,421 is
also an EGRET source (Hartman et al.\ 1999).  The VHE emission from
Mrk\,421 is extremely variable, showing flaring activity on time
scales as short as 15 minutes (Gaidos et al.\ 1996) with little or no
baseline level emission (Buckley et al.\ 1996).  The spectrum of
Mrk\,421 is consistent with a power law that extends to at least
10\,TeV with no evidence of a sharp cut-off, and no evidence of
variability (e.g., Krennrich et al.\ 1999a,b).  Multiwavelength campaigns
on Mrk\,421 (e.g., Buckley et al.\ 1996) have revealed correlations
between X-rays and VHE $\gamma$-rays and evidence for correlated
optical/UV variability.  The flux amplitude of the X-ray and VHE
$\gamma$-ray variations was similar and the variability time profiles
were the same, on day-scales.  These rapid, correlated variations have
permitted stringent limits to be placed on the Doppler factor and
magnetic fields of the Mrk\,421 jet (e.g., Buckley et al.\ 1996) and
these data have become important tests for emission models (e.g.,
Mannheim 1998; Buckley 1998; Tavecchio, Maraschi \& Ghisellini 1998).

Despite the successes of these campaigns, the light curves were not
densely sampled, so the multiwavelength variability could not be
measured on time-scales less than one day.  In order to better
measure the flaring behavior of Mrk\,421, several 
more intense multiwavelength campaigns were conducted in
1998 using longer exposures in X-rays and VHE $\gamma$-rays, and
combining the data from several VHE $\gamma$-ray telescopes.  Here, we
present the results of a campaign in 1998 April with 
{\it Beppo}SAX and the Whipple $\gamma$-ray telescope.

\begin{wrapfigure}[29]{r}{3.0in}
\vspace*{-0.4in}
\centerline{\epsfig{file=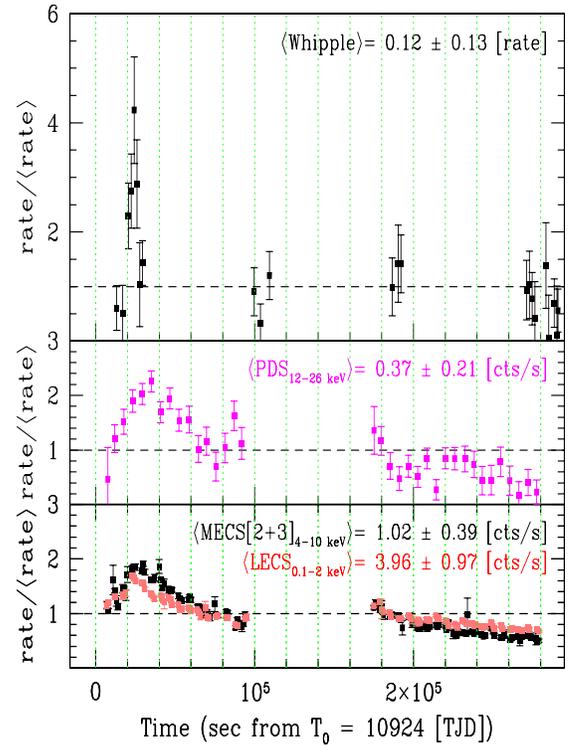,width=2.9in}}
\caption{The $\gamma$-ray and X-ray light curves of Mrk 421 from 1998
April 21 to 24 (UT).  The top panel shows data for E$>$2\,TeV.  The
middle and bottom panels show {\it Beppo}SAX data for the three
instruments with energy ranges specified in the plots.  All of the
count rates are normalized to their respective averages (listed in the
panels) during this observation period.
\label{m4lc.fig}
}
\end{wrapfigure}

\section{Observations:}
\label{observ.sec}

\subsection{BeppoSAX:}
\label{sax.sec}

The scientific payload carried by {\it Beppo}SAX is fully described in
Boella et al.\ (1997a).  The data of interest here derive from three
coaligned instruments, the Low Energy Concentrator Spectrometer (LECS,
0.1-10 keV, Parmar et al.\ 1997), the Medium Energy Concentrator
Spectrometer (MECS, 2-10 keV, Boella et al.\ 1997b) and the Phoswich
Detector System (PDS, 12-300 keV, Frontera et al.\ 1997).

The observations with {\it Beppo}SAX reported here consist of two
exposures lasting approximately 100 kiloseconds each.  The data reduction for
the PDS was done using the XAS software (Chiappetti \& Dal Fiume
1997), while for the LECS and MECS linearized cleaned event files
generated at the {\it Beppo}SAX Science Data Centre (SDC) were
used. No appreciable difference was found extracting the MECS data
with the XAS software.  Light curves were accumulated from each
instrument with the usual choices for extraction radius and background
subtraction as described in Chiappetti et al.\ (1999).

\subsection{Whipple:}
\label{whipple.sec}

The VHE $\gamma$-ray observations were made with the Whipple
Observatory 10\,m telescope (Cawley et al.\ 1990).  At the time of
these observations, the telescope camera consisted of 331
photomultiplier tubes with a combined field of view of 4.8$^\circ$.
Also, light-cones were not in place and this, as well as reduced
reflectivity of the mirrors, resulted in a somewhat higher energy
threshold than usual for the telescope, $\sim$500\,GeV.  Events were
parameterized with a standard moment analysis and candidate
$\gamma$-rays were selected using a variation of the Supercuts
analysis (Reynolds et al.\ 1993) appropriate for the large camera
field of view and for maintaining a constant energy threshold as a
function of observation elevation (see Table~\ref{norm.table} and
discussion below).

\begin{table}
\caption{Elevation dependent parameter cuts for Mrk\,421 analysis
\label{norm.table}
}
\begin{center}
\begin{tabular}{c|c|c|c|c} \hline\hline
 & \multicolumn{4}{|c}{Observation zenith angle} \\
Parameter & $<$35$^\circ$ & 40$^\circ$ & 45$^\circ$ & 55$^\circ$ \\ \hline
$\alpha$ & $< 7.5^\circ$ & $< 7.5^\circ$ & $< 7.5^\circ$ & $< 10.0^\circ$ \\
{\it length} & $>0.10^\circ$, $<0.45^\circ$ & $>0.10^\circ$, $<0.45^\circ$ 
 & $>0.10^\circ$, $<0.38^\circ$ & $>0.10^\circ$, $<0.30^\circ$ \\
{\it width} & $>0.04^\circ$, $<0.18^\circ$ & $>0.04^\circ$, $<0.15^\circ$ 
 & $>0.04^\circ$, $<0.13^\circ$ & $>0.04^\circ$, $<0.12^\circ$ \\
{\it distance} & $>0.30^\circ$, $<1.60^\circ$ & $>0.30^\circ$, $<1.60^\circ$ 
 & $>0.30^\circ$, $<1.60^\circ$ & $>0.30^\circ$, $<1.60^\circ$ \\
{\it asymmetry} & $> 0.0$ & $> 0.0$ & $> 0.0$ & $> 0.0$ \\
{\it length/size} & $\cdots$ & $\cdots$ & $\cdots$ & $< 0.00078^\circ$/d.c.$^b$ \\
{\it max2} & $> 80$\,d.c. & $> 80$\,d.c. & $> 80$\,d.c. & $> 60$\,d.c. \\ \hline
{\it size}$^c$ & $>1300$\,d.c. & $>900$\,d.c. & $>600$\,d.c. & $>200$\,d.c. \\
Effective area ratio$^d$ & 1.0 & 1.5 & 2.6 & 4.7 \\ \hline
\multicolumn{5}{l}{$^a$1\,d.c. = 1 digital count.}\\
\multicolumn{5}{l}{$^b$Size cut required for energy threshold of 2\,TeV.} \\
\multicolumn{4}{l}{$^c$Ratio is relative to the effective area at a zenith 
angle of 20$^\circ$.} \\
\end{tabular}
\end{center}
\end{table}

Observations were taken on the nights of 1998 April 21, 22, 23 and 24.
To permit longer observations within each night, data were taken over
a large range of zenith angles ($\sim$7$^\circ$ to 60$^\circ$).  The
collection area and energy threshold increase with zenith angle and
the $\gamma$-ray selection is a function of zenith angle, so the
observed $\gamma$-ray rates can change as a function of elevation even
from a source of constant $\gamma$-ray emission.  To obtain a
light-curve which shows only the intrinsic source variations, it was
necessary to determine software cuts which result in a constant energy
threshold as a function of elevation and to calculate collection areas
as a function of elevation for those energy thresholds in order to
normalize the $\gamma$-ray rates.  Because contemporaneous
observations of the Crab Nebula were not available with sufficient
statistics over all the zenith angles, the analysis presented here
relies entirely on Monte Carlo shower simulations.

Simulated $\gamma$-ray induced showers at zenith angles of 20$^\circ$,
40$^\circ$, 45$^\circ$, and 55$^\circ$ were generated with ISUSIM
(Mohanty et al.\ 1998) to determine the size cut required at each
zenith angle to obtain a common energy threshold of 2\,TeV and to
estimate the collection areas at these zenith angles for the 2\,TeV
energy threshold.  To normalize the $>$2\,TeV rate measurements at the
different elevations we multiply them by the ratio of effective
collection area at a given zenith angle to the effective collection
area at a zenith angle of 20$^\circ$.  The software trigger threshold
applied at each zenith angle to set the energy threshold at 2\,TeV and
the ratio of effective areas for the four zenith angle ranges are
shown in Table~\ref{norm.table}.  The results reported here are based
on limited statistics and are therefore preliminary.  The aim of this
analysis is to derive normalized fluxes as a function of time rather
than absolute fluxes and energy spectra.

\section{Results:}
\label{results.sec}

The light curves for the $\gamma$-ray and X-ray observations are shown
in Figure~\ref{m4lc.fig}.  Three X-ray energy bands are shown and the
$\gamma$-ray light curve shows the normalized E$>$2\,TeV data for the
measurements.  Each $\gamma$-ray point represents a 28 minute
observation.  The count rates for the measurements are normalized to
their respective averages for these observations.  The rise and fall
of a large amplitude flare is clearly evident in all data sets on the
first day of observations.  Observations after the first day did have
detectable fluxes, but showed no significant variability on day or
shorter time-scales.  For the observations on April 21, the amplitude
of the X-ray flaring increases with increasing energy, but it is close
to a factor of 2 in all three bands.  The $>$2\,TeV light curve shows
a 4-fold variation in flux.  The flux in the 0.1-2\,keV, 4-10\,keV,
and $>$2\,TeV energy bands peaks at approximately the same time
(within one-half of one hour), but the decay time for the TeV light
curve is significantly shorter than that of the LECS and MECS light
curves.  The 12-26\,keV light curve measured by the PDS instrument
appears to peak slightly later than the others, but the statistical
uncertainty in the data precludes a definitive measurement.  A
detailed investigation of possible leads or lags in the data is
underway.  The TeV spectrum does not change significantly during the
rapid flare, nor is it significantly different than previous
measurements (Krennrich et al.\ 1999b).

\section{Discussion and Conclusions:}
\label{disc.sec}

These observations show, for the first time, that the TeV and keV
fluxes from Mrk\,421 are correlated on hour time scales while at the
same time indicate that the $\gamma$-rays and X-rays are not
completely correlated.  Neither the larger variability amplitude at
TeV energies than at X-rays nor the difference in the variability time
scales have been seen previously.  The reason for the difference in
the decay time-scale of the flare at TeV and keV energies is not
clear.  The differences could reflect the nature of the flaring
mechanism.  For example, a variation in the electron spectrum and the
energy density of the low energy photons up-scattered to TeV energies
(Maraschi et al. 1999) might produce such a flare.  The differences
may also indicate that the particles which produce the X-rays are not
the same as the particles which produce the TeV $\gamma$-rays.  This
is possible in models where the progenitor particles are electrons or
protons.  In addition, the region of the broadband spectrum of
Mrk\,421 observed by {\it Beppo}SAX spans the end of the synchrotron
emission and the onset of the high energy emission (c.f., Buckley et
al.\ 1996).  As such, the X-ray emission may reflect contributions
from more than one population of source particles, regardless of the
emission mechanism.  Detailed model fitting of this data, which is
beyond the scope of this paper, is necessary to investigate these
possibilities.

%
%
%
%
%
%
\vspace{1ex}
\begin{center}
{\Large\bf References}
\end{center}
%
Boella, G., et al.\ 1997a, A\&AS, 122, 299\\
Boella, G., et al.\ 1997b, A\&AS, 122, 327\\
Buckley, J.H., et al.\ 1996, ApJ, 472, L9\\ 
Buckley, J.H.\ 1998, Science, 279, 676\\
Cawley, M.F., et al.\ 1990, Exp. Astron., 1, 173\\
Chiappetti, L., \& Dal Fiume, D. 1996, in Proc. of the 5th Workshop
``Data Analysis in Astronomy'' (Erice), ed. V. Di Ges\'u, et al., 101\\
Chiappetti, L., et al.\ 1999, ApJ, in press\\
Dermer, C.D., Schlickeiser, R., \& Mastichiadis, A. 1992, A\&A, 256, L27\\
Frontera, F., et al.\ 1997, A\&AS, 122, 357\\
Gaidos, J.A., et al.\ 1996, Nature, 383, 319\\
Hartman, R.C., et al.\ 1999, ApJS, in press\\
Krennrich, F., et al.\ 1999b, Proc. 26th ICRC (Salt Lake City), OG 2.1.02\\
Mannheim, K.\ 1993, A\&A, 269, 67\\
Mannheim, K.\ 1998, Science, 279, 684\\
Maraschi, L., Ghisellini, G., \& Celotti, A.\ 1992, ApJ, 397, L5\\
Maraschi, L., et al.\ 1999, in TeV Astrophysics of Extragalactic Sources,
ed. M. Catanese \& T.C. Weekes, Astrop. Phys., in press\\
Mohanty, G., et al.\ 1998, Astrop. Phys., 9, 15\\
Ong, R.A.\ 1998, Phys. Reports, 305, 93\\
Parmar, A.N., et al.\ 1997, A\&AS, 122, 309\\
Petry, D., et al.\ 1996, A\&A, 311, L13\\
Punch, M., et al.\ 1992, Nature, 358, 477\\
Reynolds, P.T., et al.\ 1993, ApJ, 404, 206\\
Sikora, M., Begelman, M.C., \& Rees, M.J. 1994, ApJ, 421, 153\\
Tavecchio, F., Maraschi, L., \& Ghisellini, G. 1998, ApJ, 509, 608
\end{document}